\documentclass{article}
\usepackage{graphicx}
\graphicspath{ {./figures/} }

\usepackage[utf8]{inputenc}
\usepackage{fullpage}
\usepackage{lscape}
\usepackage{longtable}

\title{Participation in TREC 2020 COVID Track Using Continuous Active Learning}

\author{
    (Jean) Xue Jun Wang\\
    University of Waterloo\\
    Waterloo, ON, Canada\\
    xj4wang@uwaterloo.ca
    \and
    Maura R Grossman\\
    University of Waterloo\\
    Waterloo, ON, Canada\\
    maura.grossman@uwaterloo.ca
    \and
    (Kevin) Seung Gyu Hyun\\
    University of Waterloo\\
    Waterloo, ON, Canada\\
    sghyun@uwaterloo.ca}
    
\date{}
    
\begin{document}

    \maketitle

    \begin{abstract}
		We describe our participation in all five rounds of the TREC 2020 COVID Track (TREC-COVID). The goal of TREC-COVID is to contribute to the response to the COVID-19 pandemic by identifying answers to many pressing questions and building infrastructure to improve search systems \cite{vabdhlrsw2020}. All five rounds of this Track challenged participants to perform a classic ad-hoc search task on the new data collection CORD-19. Our solution addressed this challenge by applying the Continuous Active Learning model (CAL) and its variations. Our results showed us to be amongst the top scoring \textit{manual} runs and we remained competitive within all categories of submissions.
	\end{abstract}
	
	\section{Introduction}
		As the spread of COVID-19 continues around the globe, researchers, clinicians, and policy makers involved with its response are constantly searching for reliable information on the virus. This presents those of us in information retrieval (IR) and text processing communities with a unique opportunity to contribute to the response to this pandemic by building infrastructure to improve search systems and to help identify answers for some of today's most pressing questions \cite{vabdhlrsw2020}.
		The task of TREC-COVID is for participants to retrieve the most relevant documents from the CORD-19 data-set for a given set of topics.
	    To address this challenge, we implemented a system based on CAL, following the work of Grossman and Cormack in \cite{cg2016, gc2016}, using the tool kit provided as part of the Baseline Model Implementation (BMI), created by Roegiest and Cormack in \cite{rc2015}, and ourselves as the human assessors.
	
	\section{Related Work}
	    In this section, we discuss prior research on CAL. We then discuss prior research on BMI, which provides the tool kits we heavily relied upon for this challenge.
	
	    \paragraph{Continuous Active Learning (CAL).}
    	CAL is a method for finding virtually all relevant information on a particular subject within a vast sea of electronically stored information (ESI): it repeatedly refines its understanding about which of the remaining documents are most likely to be of interest, based on the users' feedback regarding the documents already judged \cite{gc2016}.
	    This protocol is most famously used in technology-assisted review (TAR) for electronic discovery in legal matters, achieving the best results reported in scientific literature to date \cite{cg2014}. Building on the CAL protocol, many implementations, such as BMI, have been highly successful at performing ad-hoc retrieval tasks, such as in the TREC 2015/2016 Total Recall Track \cite{rccg2015, gcr2016} and the TREC 2019 Decision Track \cite{absd2019}.
	
	    \paragraph{Baseline Model Implementation (BMI).}
		BMI is an augmented version of CAL. It is autonomous and was initially made available to participants of the TREC 2015/2016 Total Recall Tracks \cite{rccg2015, gcr2016}, as well as the TREC 2019 Decision Track \cite{absd2019} to provide a baseline for comparison. However, BMI turned out to be highly competitive, with none of the manual participants achieving consistently superior results to this fully automated method \cite{rc2015}.
        \\
        \\
		While BMI has been shown to generally outperform human-in-the-loop CAL implementations \cite{rc2015}, it requires labelled data, which was very limited, if available at all, for TREC-COVID; thus, we chose to insert a human back into the loop to make judgements. All other components, such as creating feature vectors, the learner, etc., were taken directly from the BMI tool kits.
		
	\section{System Overview \& General Approach}
	
    	\paragraph{Document Set Processing.} The document set used in the TREC-COVID Challenge is the COVID-19 Open Research Data-set (CORD-19). Our team opted to judge a document's relevancy using strictly the information available in the metadata file (year, authors, publisher, title, abstract) based on the work of Zhang et al. \cite{zagscg2018} which show that participants achieve higher recall using CAL when presented with only a single short excerpt rather than an entire document.
        
        \paragraph{CAL.} The following shows an outline of our specific implementation of CAL.
        \begin{itemize}
		    \item \textit{\textbf{STEP 1:}} \textit{Create a hypothetical relevant document, known as a synthetic document.}\\
		    To create the synthetic documents, we concatenated the query, question, and narrative components of the topics file provided by TREC-COVID, as shown in Figures \ref{fig:syntethic_file} and \ref{fig:xml_topics}.
    		\begin{figure}[htbp]
              \centering
              \begin{minipage}[b]{.45\textwidth}
                \includegraphics[width=1\columnwidth]{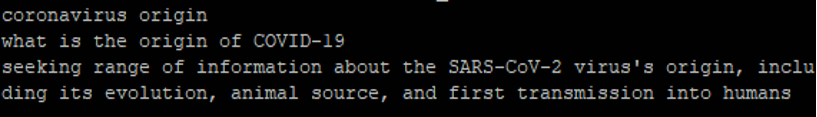}
    			\caption{The synthetic document for topic 1.}
    			\label{fig:syntethic_file}
              \end{minipage}
              \hfill
              \begin{minipage}[b]{.45\textwidth}
                \includegraphics[width=1\columnwidth]{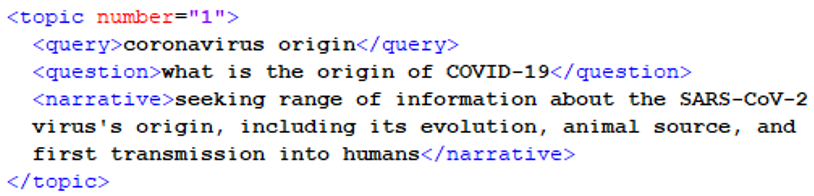}
    			\caption{Snippet of topic 1 in the xml topics file provided by TREC-COVID.}
    			\label{fig:xml_topics}
              \end{minipage}
            \end{figure}
    		
    		\item \textbf{STEP 2:} \textit{Use a machine-learning algorithm to suggest the next most-likely relevant document.}\\
    		The machine-learning algorithm we chose is Sofia-ML which Roegiest and Cormack used in their participation in the 2015 Total Recall Track \cite{rc2015}.
    		
    		\item \textbf{STEP 3:} \textit{Review the suggested documents and provide relevance feedback to the learning algorithm, indicating whether each suggested document is actually relevant or not.}\\
    		To do this, we sorted the results given by Sofia-ML in decreasing order of confidence, presenting the top most result to the human assessor using a text based user interface.
    		The judgement made by our human assessors is one of \{0-not relevant, 1-partially relevant, 2-relevant\}. This corresponds to the annotations made by biomedical experts as part of TREC-COVID following each round. As Sofia-ML does not distinguish between \textit{relevant} judgements and \textit{partially relevant} judgments, both were designated to be \textit{relevant} in training.
    		
    		\item \textbf{STEP 4:} \textit{Repeat Step 2 and 3 until very few, if any, of the suggested documents are relevant.}\\
    		Using the same stopping condition as in \cite{gcr2016}, we aimed to stop when the following criterion was met:
    		$$n \ge a * m + b$$
    		where \textit{m} is the number of relevant documents reviewed, \textit{n} is the number of irrelevant documents reviewed, \textit{a} is a constant which determines how many non-relevant documents are to be reviewed in the course of finding each relevant document, and \textit{b} is a constant which represents a fixed overhead for the number of irrelevant documents that must be reviewed.
    	\end{itemize}
    	
    	\paragraph{S-CAL.}
    	One of the major drawbacks of the CAL method outlined above is the impractical number of documents that must be reviewed when the number of relevant documents is large.
        Scalable Continuous Active Learning (S-CAL) \cite{cg2016} addresses this issue by
        \begin{enumerate}
    		\item Segmenting the corpus into batches and allowing assessors to label only a small finite sample of documents from each successive batch.
    		\item Temporarily augmenting each training set by adding a set of 100 random documents from the corpus - which is, with high probability, \textit{not relevant} for a large corpus - labelled \textit{not relevant}.
    	\end{enumerate}
        However, the stopping condition for S-CAL outlined in \cite{cg2016} is still infeasible to achieve with CORD-19 and our team size; thus, we exchange the initial dynamic stopping condition for a static goal of assessing 300 documents per topic.
    	
    	\paragraph{Hyper-parameter Tuning.}
    	Given the availability of labelled data after the first round, we performed hyper-parameter tuning on both the \textit{loop\_type} and the \textit{lambda} value to better fit CORD-19.
    	Finding no significant differences in our tests, we decided to continue with our initial values taken from \cite{rc2015}, which were decided upon discussion with the author of Sofia-ML as well as their internal experiments.
    	
\begin{figure}[h]
    		\centering
    		\includegraphics[width=.5\columnwidth]{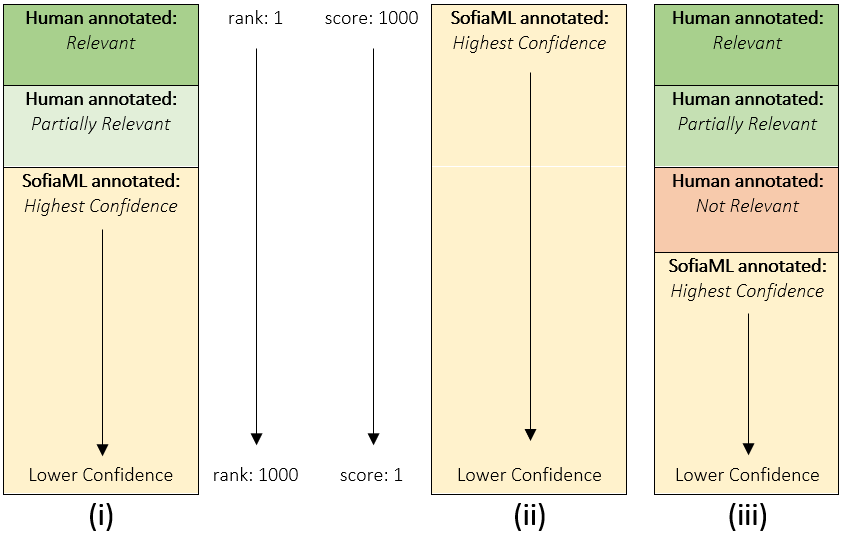}
    		\caption{Document orderings for runs.\\
    		(i) Lead with documents labelled relevant, followed by partially relevant, and finally filled by Sofia-ML's labelling of unseen documents in descending order of confidence.\\
    	    (ii) All documents are arranged using Sofia-ML. No special consideration is given to documents already assessed by a human assessor.\\
    	    (iii) Keep documents that annotators have labelled to be not relevant in the final run.}
    		\label{fig:listordering_final}
    	\end{figure}

    	\paragraph{Creating Runs.}
        To generate the results for our runs, we created lists of 1000 documents ordered as shown in Figure \ref{fig:listordering_final}.
    	
\begin{figure}[htbp]
          \centering
          \begin{minipage}[b]{.45\textwidth}
            \includegraphics[width=1\columnwidth]{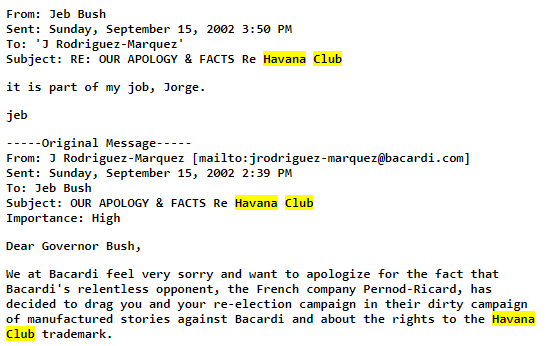}
    	    \caption{A sample document retrieved from Grossman and Cormack's online CAL platform,  https://cormack.uwaterloo.ca/cal/, showing key-term highlighting.}
    	    \label{fig:CAL_highlight}
          \end{minipage}
          \hfill
          \begin{minipage}[b]{.45\textwidth}
            \includegraphics[width=1\columnwidth]{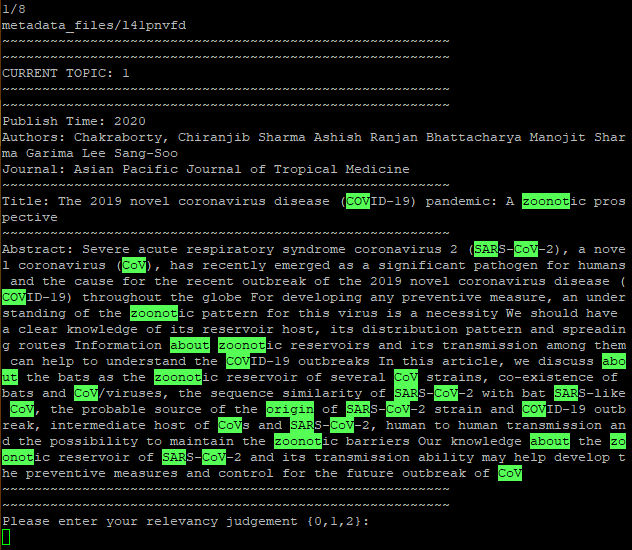}
    	    \caption{Text based user interface showing key-term highlighting.}
    	    \label{fig:UI_highlight}
          \end{minipage}
        \end{figure}

		\paragraph{Key-Term Highlighting.}
    	Key-term highlighting is a feature commonly provided by IR systems, such as Google, to assist human readers in processing information. Following the online sample of CAL, as show in Figure \ref{fig:CAL_highlight}, given as a supplement to \cite{gc2016}, we chose to highlight the top five highest-scoring words from a document, according to Sofia-ML, in our UI for assessors, as show in Figure \ref{fig:UI_highlight}.
		
	\section{Results and Discussion}
    Table \ref{tab:participation} shows the specifications of our system for each round of TREC-COVID and Table \ref{tab:results} shows our results. From these, we are able to make some interesting observations:
	\begin{enumerate}
		\item Despite our human assessors having provided more labelled documents in round 2 than round 1, our performance decreased. One possible explanation could be that, through the use of the key-term highlighting feature, our human assessor(s) exchanged quantity for quality resulting in an overall poorer model.
		
		\item Despite being able to provide more labelled documents in round 5 than 4, our performance once again decreased. One possible explanation could be that we did not perform the necessary quality control required for additional human assessors - once again, exchanging quantity for quality labels, resulting in an overall poorer performance.
		
		\item The runs ordered by method Figure \ref{fig:listordering_final}(iii) consistently outperformed our other runs. This could imply that the documents judged to be \textit{not-relevant} by our assessors are still more relevant than Sofia-ML's labelling of unseen documents.
	\end{enumerate}
    
    \begin{table}[]
        \centering
        \resizebox{\textwidth}{!}{%
        \begin{tabular}{|c|l|l|l|}
        \hline
         & \multicolumn{1}{c|}{\textbf{System Overview}} & \multicolumn{1}{c|}{\textbf{Run submissions}} & \multicolumn{1}{c|}{\textbf{System Issues}} \\ \hline
        \textbf{Round 1} & \begin{tabular}[c]{@{}l@{}}Document set processing,\\ CAL,\\ 1 assessor\end{tabular} & \begin{tabular}[c]{@{}l@{}}xj4wang\_run1: \\ ordered by method (i)\end{tabular} & \begin{tabular}[c]{@{}l@{}}Being pressed for time, we were unable to reach our stopping\\ condition, prematurely stopping after 40 document assessments\\ for each topic.\\  \\ Using sort -rn instead of sort -rg resulting in documents with \\ exponentially low confidence being sorted to the top during \\ both the assessing process and the run creation.\end{tabular} \\ \hline
        \textbf{Round 2} & \begin{tabular}[c]{@{}l@{}}Same as Round 1,\\ + Key-term highlighting\end{tabular} & \begin{tabular}[c]{@{}l@{}}xj4wang\_run3: \\ ordered by method (i)\end{tabular} & \begin{tabular}[c]{@{}l@{}}Being pressed for time, we were unable to reach our stopping \\ condition, prematurely stopping after 60 document \\ assessments for each topic.\end{tabular} \\ \hline
        \textbf{Round 3} & \begin{tabular}[c]{@{}l@{}}Same as Round 2,\\ $\pm$ Switching out CAL for S-CAL,\\ + 1 additional assessor, total of 2\end{tabular} & \begin{tabular}[c]{@{}l@{}}xj4wang\_run1: \\ ordered by method (iii)\\ \\ xj4wang\_run2: \\ ordered by method (ii)\\ \\ xj4wang\_run3: \\ ordered by method (i)\end{tabular} & \begin{tabular}[c]{@{}l@{}}Being pressed for time, we were unable to reach our stopping \\ condition for every topic.\end{tabular} \\ \hline
        \textbf{Round 4} & \begin{tabular}[c]{@{}l@{}}Same as Round 3,\\ + 1 additional assessor, total of 3\end{tabular} & Same as Round 3 &  \\ \hline
        \textbf{Round 5} & \begin{tabular}[c]{@{}l@{}}Same as Round 4,\\ + 2 additional assessor, total of 5\end{tabular} & Same as Round 3 &  \\ \hline
        \end{tabular}%
        }
        \caption{Specifications of system design for each round of TREC-COVID.}
        \label{tab:participation}
        \end{table}
    
    \renewcommand{\arraystretch}{1.5} 
    
    \begin{table}[]
\centering
\resizebox{\textwidth}{!}{%
\begin{tabular}{|c|c|c|c|c|c|c|c|c|c|c|c|}
\hline
 & \textbf{ROUND1} & \textbf{ROUND2} & \multicolumn{3}{c|}{\textbf{ROUND3}} & \multicolumn{3}{c|}{\textbf{ROUND4}} & \multicolumn{3}{c|}{\textbf{ROUND5}} \\ \hline
 & \textbf{\begin{tabular}[c]{@{}c@{}}xj4wang\\ \_run1\end{tabular}} & \textbf{\begin{tabular}[c]{@{}c@{}}xj4wang\\ \_run3\end{tabular}} & \textbf{\begin{tabular}[c]{@{}c@{}}xj4wang\\ \_run1\end{tabular}} & \textbf{\begin{tabular}[c]{@{}c@{}}xj4wang\\ \_run2\end{tabular}} & \textbf{\begin{tabular}[c]{@{}c@{}}xj4wang\\ \_run3\end{tabular}} & \textbf{\begin{tabular}[c]{@{}c@{}}xj4wang\\ \_run1\end{tabular}} & \textbf{\begin{tabular}[c]{@{}c@{}}xj4wang\\ \_run2\end{tabular}} & \textbf{\begin{tabular}[c]{@{}c@{}}xj4wang\\ \_run3\end{tabular}} & \textbf{\begin{tabular}[c]{@{}c@{}}xj4wang\\ \_run1\end{tabular}} & \textbf{\begin{tabular}[c]{@{}c@{}}xj4wang\\ \_run2\end{tabular}} & \textbf{\begin{tabular}[c]{@{}c@{}}xj4wang\\ \_run3\end{tabular}} \\ \hline
\textbf{Number of topics} & 30 & 35 & \multicolumn{3}{c|}{40} & \multicolumn{3}{c|}{45} & \multicolumn{3}{c|}{50} \\ \hline
\textbf{Total number retrieved} & 30,000 & 35000 & 39938 & 39941 & 39942 & 41931 & 42160 & 42241 & 49923 & 49927 & 49928 \\ \hline
\textbf{Total relevant} & 2352 & 3002 & \multicolumn{3}{c|}{4698} & \multicolumn{3}{c|}{5824} & \multicolumn{3}{c|}{10910} \\ \hline
\textbf{Total relevant retrieved} & 1216 & 1768 & 2857 & 2794 & 2742 & 3241 & 3024 & 2950 & 6188 & 5889 & 5743 \\ \hline
\textbf{MAP} & 0.2367 & 0.2210 & 0.2836 & 0.2534 & 0.2751 & 0.2963 & 0.2774 & 0.2775 & 0.2647 & 0.2509 & 0.2448 \\ \hline
\textbf{Mean Bpref} & 0.4599 & 0.4823 & 0.5681 & 0.5537 & 0.5464 & 0.5507 & 0.5216 & 0.5084 & 0.5254 & 0.5062 & 0.4912 \\ \hline
\textbf{Mean NDCG@10} & 0.6513 & 0.5907 & 0.7431 & 0.6252 & 0.7413 &  &  &  &  &  &  \\ \hline
\textbf{Mean NDCG@20} &  &  &  &  &  & 0.7019 & 0.6855 & 0.7019 & 0.6663 & 0.6685 & 0.6663 \\ \hline
\textbf{Mean RBP(p=0.5)} &  & \begin{tabular}[c]{@{}c@{}}0.6546\\ +0.0041\end{tabular} & \begin{tabular}[c]{@{}c@{}}0.7300\\ +0.0407\end{tabular} & \begin{tabular}[c]{@{}c@{}}0.6303\\ +0.2002\end{tabular} & \begin{tabular}[c]{@{}c@{}}0.7299\\ +0.0412\end{tabular} & \begin{tabular}[c]{@{}c@{}}0.7946\\ +0.0194\end{tabular} & \begin{tabular}[c]{@{}c@{}}0.7486\\ +0.0201\end{tabular} & \begin{tabular}[c]{@{}c@{}}0.7946\\ +0.0194\end{tabular} & \begin{tabular}[c]{@{}c@{}}0.7767\\ +0.0018\end{tabular} & \begin{tabular}[c]{@{}c@{}}0.7539\\ +0.0015\end{tabular} & \begin{tabular}[c]{@{}c@{}}0.7767\\ +0.0018\end{tabular} \\ \hline
\textbf{P@5} & 0.8333 & 0.7314 & 0.8350 & 0.7000 & 0.8350 & 0.8933 & 0.8844 & 0.8933 & 0.8640 & 0.8440 & 0.8640 \\ \hline
\textbf{P@10} & 0.7167 & 0.6400 & 0.8325 & 0.7050 & 0.8275 & 0.8422 & 0.8400 & 0.8422 & 0.8080 & 0.8080 & 0.8080 \\ \hline
\textbf{P@15} & 0.6067 & 0.5505 & 0.7200 & 0.6733 & 0.7183 & 0.7644 & 0.7719 & 0.7644 & 0.7307 & 0.7427 & 0.7307 \\ \hline
\textbf{P@20} & 0.5417 & 0.4957 & 0.6625 & 0.6400 & 0.6588 & 0.7244 & 0.7278 & 0.7244 & 0.6780 & 0.7030 & 0.6780 \\ \hline
\textbf{P@30} & 0.4522 & 0.4162 & 0.5633 & 0.5517 & 0.5608 & 0.6378 & 0.6326 & 0.6385 & 0.6247 & 0.6400 & 0.6247 \\ \hline
\textbf{R-Precision} & 0.2843 & 0.2644 & 0.3325 & 0.3111 & 0.3280 & 0.3503 & 0.3266 & 0.3316 & 0.3207 & 0.3076 & 0.2984 \\ \hline
\end{tabular}%
}
\caption{Results of all runs of team xj4wang.}
\label{tab:results}
\end{table}
    
    \begin{table}[]
\centering
\resizebox{\textwidth}{!}{%
\begin{tabular}{cccccccccc}
\hline
\multicolumn{1}{|c|}{} & \multicolumn{3}{c|}{\textbf{ROUND1}} & \multicolumn{3}{c|}{\textbf{ROUND2}} & \multicolumn{3}{c|}{\textbf{ROUND3}} \\ \hline
\multicolumn{1}{|c|}{} & \multicolumn{1}{c|}{\textbf{\begin{tabular}[c]{@{}c@{}}xj4wang\\ \_run1\end{tabular}}} & \multicolumn{1}{c|}{\textbf{\begin{tabular}[c]{@{}c@{}}Highest \\ manual \\ run score\end{tabular}}} & \multicolumn{1}{c|}{\textbf{\begin{tabular}[c]{@{}c@{}}Highest \\ overall \\ run score\end{tabular}}} & \multicolumn{1}{c|}{\textbf{\begin{tabular}[c]{@{}c@{}}xj4wang\\ \_run3\end{tabular}}} & \multicolumn{1}{c|}{\textbf{\begin{tabular}[c]{@{}c@{}}Highest \\ manual \\ run score\end{tabular}}} & \multicolumn{1}{c|}{\textbf{\begin{tabular}[c]{@{}c@{}}Highest \\ overall \\ run score\end{tabular}}} & \multicolumn{1}{c|}{\textbf{\begin{tabular}[c]{@{}c@{}}xj4wang\\ \_run1\end{tabular}}} & \multicolumn{1}{c|}{\textbf{\begin{tabular}[c]{@{}c@{}}Highest \\ manual \\ run score\end{tabular}}} & \multicolumn{1}{c|}{\textbf{\begin{tabular}[c]{@{}c@{}}Highest \\ overall \\ run score\end{tabular}}} \\ \hline
\multicolumn{1}{|c|}{\textbf{MAP}} & \multicolumn{1}{c|}{0.2367} & \multicolumn{1}{c|}{0.3008} & \multicolumn{1}{c|}{0.3128} & \multicolumn{1}{c|}{0.2210} & \multicolumn{1}{c|}{0.338} & \multicolumn{1}{c|}{0.338} & \multicolumn{1}{c|}{0.2836} & \multicolumn{1}{c|}{0.3244} & \multicolumn{1}{c|}{0.3333} \\ \hline
\multicolumn{1}{|c|}{\textbf{Mean Bpref}} & \multicolumn{1}{c|}{0.4599} & \multicolumn{1}{c|}{0.5294} & \multicolumn{1}{c|}{0.5294} & \multicolumn{1}{c|}{0.4823} & \multicolumn{1}{c|}{0.5679} & \multicolumn{1}{c|}{0.5679} & \multicolumn{1}{c|}{0.5681} & \multicolumn{1}{c|}{0.5828} & \multicolumn{1}{c|}{0.6084} \\ \hline
\multicolumn{1}{|c|}{\textbf{Mean NDCG@10}} & \multicolumn{1}{c|}{0.6513} & \multicolumn{1}{c|}{0.6844} & \multicolumn{1}{c|}{0.6844} & \multicolumn{1}{c|}{0.5907} & \multicolumn{1}{c|}{0.6893} & \multicolumn{1}{c|}{0.6893} & \multicolumn{1}{c|}{0.7431} & \multicolumn{1}{c|}{0.7431} & \multicolumn{1}{c|}{0.7740} \\ \hline
\multicolumn{1}{|c|}{\textbf{Mean RBP(p=0.5)}} & \multicolumn{1}{c|}{0.724} & \multicolumn{1}{c|}{0.7699} & \multicolumn{1}{c|}{0.7699} & \multicolumn{1}{c|}{0.6546} & \multicolumn{1}{c|}{0.7547} & \multicolumn{1}{c|}{0.7547} & \multicolumn{1}{c|}{0.7300} & \multicolumn{1}{c|}{0.7770} & \multicolumn{1}{c|}{0.8068} \\ \hline
\multicolumn{1}{|c|}{\textbf{P@5}} & \multicolumn{1}{c|}{0.8333} & \multicolumn{1}{c|}{0.8333} & \multicolumn{1}{c|}{0.8333} & \multicolumn{1}{c|}{0.7314} & \multicolumn{1}{c|}{0.8514} & \multicolumn{1}{c|}{0.8514} & \multicolumn{1}{c|}{0.8350} & \multicolumn{1}{c|}{0.8350} & \multicolumn{1}{c|}{0.8950} \\ \hline
\textbf{} &  &  &  &  &  &  &  &  &  \\ \cline{1-8}
\multicolumn{1}{|c|}{} & \multicolumn{3}{c|}{\textbf{ROUND4}} & \multicolumn{4}{c|}{\textbf{ROUND5}} & \textbf{} & \textbf{} \\ \cline{1-8}
\multicolumn{1}{|c|}{} & \multicolumn{1}{c|}{\textbf{\begin{tabular}[c]{@{}c@{}}xj4wang\\ \_run1\end{tabular}}} & \multicolumn{1}{c|}{\textbf{\begin{tabular}[c]{@{}c@{}}Highest \\ manual \\ run score\end{tabular}}} & \multicolumn{1}{c|}{\textbf{\begin{tabular}[c]{@{}c@{}}Highest \\ overall \\ run score\end{tabular}}} & \multicolumn{1}{c|}{\textbf{\begin{tabular}[c]{@{}c@{}}xj4wang\\ \_run1\end{tabular}}} & \multicolumn{1}{c|}{\textbf{\begin{tabular}[c]{@{}c@{}}xj4wang\\ \_run2\end{tabular}}} & \multicolumn{1}{c|}{\textbf{\begin{tabular}[c]{@{}c@{}}Highest \\ manual \\ run score\end{tabular}}} & \multicolumn{1}{c|}{\textbf{\begin{tabular}[c]{@{}c@{}}Highest \\ overall \\ run score\end{tabular}}} & \textbf{} & \textbf{} \\ \cline{1-8}
\multicolumn{1}{|c|}{\textbf{MAP}} & \multicolumn{1}{c|}{0.2963} & \multicolumn{1}{c|}{0.3923} & \multicolumn{1}{c|}{0.4681} & \multicolumn{1}{c|}{0.2647} & \multicolumn{1}{c|}{0.2509} & \multicolumn{1}{c|}{0.3254} & \multicolumn{1}{c|}{0.4731} &  &  \\ \cline{1-8}
\multicolumn{1}{|c|}{\textbf{Mean Bpref}} & \multicolumn{1}{c|}{0.5507} & \multicolumn{1}{c|}{0.6317} & \multicolumn{1}{c|}{0.6801} & \multicolumn{1}{c|}{0.5254} & \multicolumn{1}{c|}{0.5062} & \multicolumn{1}{c|}{0.5255} & \multicolumn{1}{c|}{0.6378} &  &  \\ \cline{1-8}
\multicolumn{1}{|c|}{\textbf{Mean NDCG@20}} & \multicolumn{1}{c|}{0.7019} & \multicolumn{1}{c|}{0.7019} & \multicolumn{1}{c|}{0.7843} & \multicolumn{1}{c|}{0.6663} & \multicolumn{1}{c|}{0.6685} & \multicolumn{1}{c|}{0.6877} & \multicolumn{1}{c|}{0.8496} &  &  \\ \cline{1-8}
\multicolumn{1}{|c|}{\textbf{Mean RBP(p=0.5)}} & \multicolumn{1}{c|}{0.7946} & \multicolumn{1}{c|}{0.8056} & \multicolumn{1}{c|}{0.8838} & \multicolumn{1}{c|}{0.7767} & \multicolumn{1}{c|}{0.7539} & \multicolumn{1}{c|}{0.7789} & \multicolumn{1}{c|}{0.9399} &  &  \\ \cline{1-8}
\multicolumn{1}{|c|}{\textbf{P@20}} & \multicolumn{1}{c|}{0.7244} & \multicolumn{1}{c|}{0.7278} & \multicolumn{1}{c|}{0.8211} & \multicolumn{1}{c|}{0.6780} & \multicolumn{1}{c|}{0.7030} & \multicolumn{1}{c|}{0.74} & \multicolumn{1}{c|}{0.8760} &  &  \\ \cline{1-8}
\end{tabular}%
}
\caption{Results of team xj4wang and the maximum scores obtained per measurement across all different teams.}
\label{tab:results_max}
\end{table}
	
	\section{Conclusion}
	
	In this paper, we report on our participation to the TREC 2020 COVID Track rounds 1 though 5, describing our approach, results, and lessons learned.
	We initially use CAL \cite{gc2016}, implemented using tools from BMI's feature kit \cite{rc2015}, with ourselves as the annotators.
	The large human labelling effort required for our system motivated us to implement a key-term highlighting feature, use S-CAL \cite{cg2016}, and recruit more human assessors.
	The results in Table \ref{tab:results_max} show us to be among the top-scoring \textit{manual} runs and competitive within all categories of submissions throughout all rounds.
	Our results in Table \ref{tab:results} also bring up an age-old question of quantity versus quality when it comes to data in IR. 
	
	\section*{Acknowledgement}
	
	A special thanks goes to Gordon Cormack for his valuable guidance and Anmol Singh for his insights. 
	\\
	We would also like to thank Charlotte Stinson, Eric Sheen, and Solaiappan Alagappan for the time and effort they spent assessing these documents.
	\\
	\\
	This research was funded in part by a Natural Sciences and Engineering Research Council of Canada (NSERC) Discovery Grant awarded to Maura R. Grossman, No. RGPIN-2017-04239, titled “Evaluation of High-Recall Human-in-the-Loop Information Retrieval Technology.”

	\bibliographystyle{plain}
	\bibliography{bib}

\end{document}